# A Systematic Review on Human Modeling: Digging into Human Digital Twin Implementations

Heribert Pascual*, Xavi Masip-Bruin*, Albert Alonso**, Judit Cerdá*
*Advanced Network Architectures Lab (CRAAX), UPC Barcelona Tech
**Fundació Clínic per a la Recerca Biomèdica (FCRB), Barcelona
{heribert.pascual, xavier.masip, judit.cerda}@upc.edu; aalonso@recerca.clinic.cat

*Abstract* — **Human Digital Twins (HDTs) are digital replicas of humans that either mirror a complete human body, some parts of it –as can be organs, flows, cells–, or even human behaviors. An HDT is a human specific replica application inferred from the digital twin (DT) manufacturing concept, defined as a technique that creates digital replicas of physical systems or processes aimed at optimizing their performance and supporting more accurate decision-making processes. The main goal of this paper is to provide readers with a comprehensive overview of current efforts in the HDT field, by browsing its basic concepts, differences with DTs, existing developments, and the distinct areas of application. The review methodology includes an exhaustive review of scientific literature, patents, and industrial initiatives, as well as a discussion about ongoing and foreseen HDT research activity, emphasizing its potential benefits and limitations.**

## I. Introduction and Motivation

Human digital twins (HDTs) are digital replicas of either a human body or some of its organs, used in many distinct sectors and verticals, including health (e.g. monitoring diseases progression), sport (e.g. supervising athlete's performance), or manufacturing (e.g. scheduling tasks in factories), just to name a few. An HDT is an extension of the digital twin (DT) concept, standing for the creation of digital replicas of physical systems or processes in order to both optimizing their performance and improving the decision-making processes. It is largely known the fact that nowadays the concept of DTs is gaining popularity, thus being applied in a wide range of sectors.

### A. Motivation

The main goal of this survey is to provide readers with a comprehensive overview of current HDT efforts, particularly focusing on the following objectives:

*1) To understand the basic concepts of a DT, including its definition, types, life cycle, and characteristics, thus easing readers to acquire the necessary background about DTs, particularly including their capabilities and limitations, making it easier for them to contextualize HDTs within the broader field of DT.*

*2) To introduce the HDT concept and explain the key differences vs a DT, also illustrating how a HDT may be used.*

*3) To dig into the scientific literature, patents, and industrial initiatives for current efforts addressing HDT implementations, thus providing an extensive overview of real efforts towards developing HDTs.*

*4) To identify the main sectors HDTs are being and will be applied to, thus emphasizing potential HDTs applicability as well as how different sectors may benefit from its deployment.*

*5) To analyze the rationales for HDTs to be used in all envisioned specific verticals, emphasizing specific goals, challenges and research objectives, while also highlighting potential benefits and limitations..*

### B. The Digital Twin Concept (DT)

A DT is a virtual replica of a physical system, process, or product, that is periodically updated with data collected from its corresponding physical entity and environment. By replicating the behavior of the physical entity under different conditions and analyzing the results, DTs can be used to optimize performance, prevent hardware failures, anticipate maintenance needs, and simulate processes by sending the actuations over its physical entity. The concept of DTs has been widely studied, as summarized in section 2A.

### C. The Human Digital Twin Concept (HDT)

HDTs apply the DT's replica concept to humans, thus creating digital replicas of either a human or part of it (e.g., organ). HDTs are created by first collecting data from various sources, such as cameras, sensors, wearables, medical devices, medical records, etc., and then using these data to build a digital model. As is later explained in this paper, this information may be processed using different technologies, e.g., Artificial Intelligence (AI), artificial vision, 3D simulations, rule based, etc. The rationale behind the DT concept makes it easier to assess that HDTs may be used to deploy proactive strategies that, in the case of humans, may notably contribute to several aspects, e.g. predict a disease behavior anticipate the disease status or performance of humans under several conditions and contexts (including for example healthcare, manufacturing, and sports), proactively support patient prognosis, etc.

Indeed, it is widely accepted that HDTs may have the potential to bring many notable benefits to humans, such as improving patients care, optimizing manufacturing processes, enhancing safety in transportation systems, or optimizing athletic performance, to name a few. For example, in health, HDTs can be used to predict the likelihood of a patient to develop a certain condition, or to evaluate the effectiveness of different treatment options. In manufacturing, HDTs can be used to optimize production schedules, thus improving quality. In transportation, HDTs can be used to optimize vehicles design, improve driver safety systems, and reduce accidents. Finally, in sports, HDTs can be used to evaluate training programs, optimize nutrition and hydration, and improve performance.

### *D. Review Methodology*

A search was conducted in September 2022 in Google Scholar for the first 100 results ordered by relevance for the terms "human AND digital AND twin*". The same search was also conducted in the Web of Science search engine. In this case 200 results were collected (half of them were patents), in order to analyze a similar volume of results on both searches. The results of both searches (different than patents), were put together in a list, from which repeated papers were eliminated. Only papers written in English were revised by two reviewers, and those that did not include implementation of HDT were rejected. Papers that were marked as not having any implementation by both reviewers were discarded. Papers positively marked by at least one reviewer were included in the eligibility phase. During the eligibility phase, each paper was revised in depth and given a score by the two reviewers, based on the proposal's feasibility, the implementation, and the overall proposal quality. The scores given by the reviewers could be zero, one, or two. Finally, only the most relevant and high-quality papers, as determined by a final score equal or greater than three, were included in the survey.

Patents that did not contain an English abstract were excluded from further analysis. The remaining patents abstracts were evaluated by two reviewers to detect an HDT implementation. Each reviewer assigned a score of 1 if an implementation was identified, and a score of 0 if it was not. To include the patents that scored only 1, a third reviewer was consulted to decide to give the last point. Only patents that received a score of 2 were included in the summary.

Additionally, we undertook many different searches in some search engines looking for private companies' implementations, that allowed us to retrieve documents from grey literature. These private companies' implementations were also considered, as constituting part of the state-of-the art in HDT technology.

### *E. Related Surveys*

Previous work has already been done to summarize the efforts in the DT area, turning into several surveys already published. These surveys are also revisited and analyzed in this paper. Particularly, surveys [1], [2], [3],[4], [5] providing a comprehensive overview of the DTs basic concepts from a global perspective, and surveys conducted on the DT topic [6], [7], [8] are interesting for further research in other areas of HDT, not just the implementation like, were considered.

## II. DIGITALTWIN BASIC CONCEPTS

The concept of a DT was first introduced by Dr. Michael Grieves in a course on Product Lifecycle Management at the University of Michigan, and later published in a whitepaper [9]. In 2010, NASA researchers published Grieves' idea in their roadmap [10] and began investigating ways to use DTs to reduce costs and resources in their space assets. Recently, the interest in DTs has grown significantly mainly motivated by current advances in network capabilities, data collection and delivery technologies (leveraging concepts such as the Internet of Things, big data, and industry 4.0), as well as the availability of data.

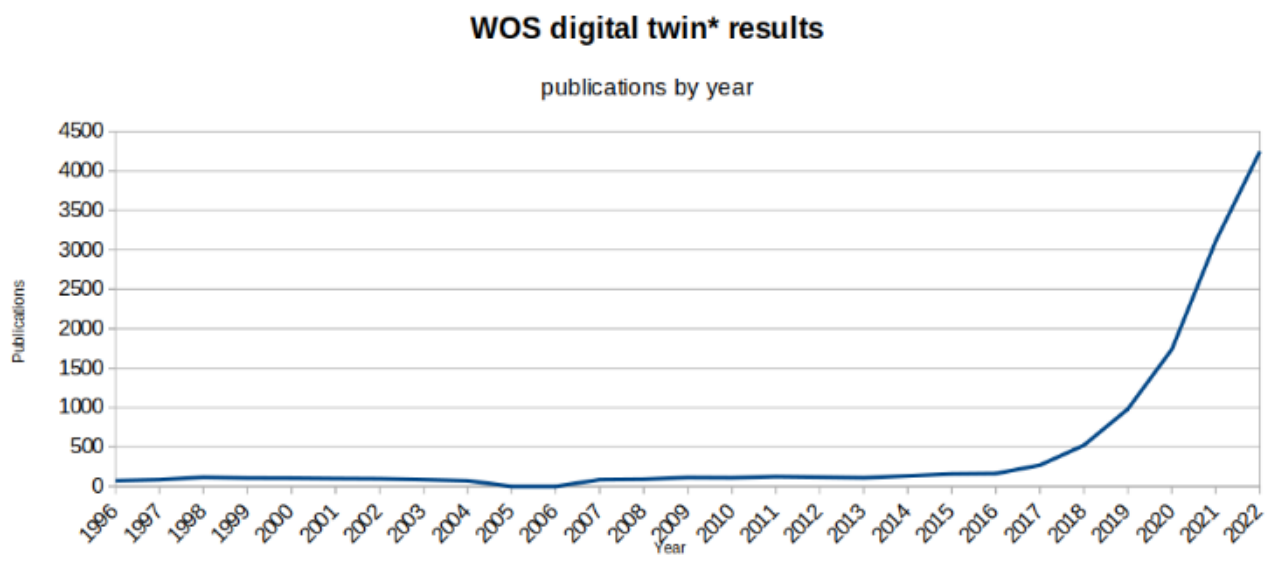

Figure 1 WOS DT publications by year

In this context, DTs have become an important tool for simulating and analyzing the behavior and characteristics of physical entities. Some of the key features of DTs include their ability to:

*1) Mimic the structure, environment, and behavior of physical entities.*

*2) Be dynamically updated with data from the physical entity throughout its life cycle. DTs supporting near to real time updates are called active DTs, while semi-active DTs only support asynchronous updates.*

*3) Perform simulations, detect failures, and make predictions about the performance of physical entities*

*4) Transfer the outcome of simulations and analytics to the physical entity in order to improving its performance by both preventing hardware failures and/or malfunctions, and anticipating maintenance needs*

In the following subsections, we will revise the definition, key features, life cycle, and characteristics of DTs, also revisiting some currently common uses for this technology, intended to provide the reader with a more thorough understanding of the concept and its key points.

### *A. Definition*

There is no a universally agreed definition of a DT within the scientific community, in the terms of what we may call a standard. However, several definitions may be found, customized to suit specific sectors and verticals. For example, the American Institute of Aeronautics and Astronautics (AIAA) [11], defines a DT as "*a set of virtual information constructs that mimics the structure, context, and behavior of an individual/unique physical asset, or a group of physical assets, is dynamically updated with data from its physical twin throughout its life cycle and informs decisions that realize value*". A slightly distinct approach by Rainer Stark and Thomas Damerau [12], defines a DT as "*A digital representation of an active unique product (real device, object, machine, service, or intangible asset) or unique product-service system (a system consisting of a product and a related service) that comprises its selected characteristics, properties, conditions, and behaviors by means of models, information, and data within a single or even across multiple life cycle phases*". Many other DT definitions exists [13], and even other similar concepts can be also found in the literature, such as digital mirroring and digital shadow, which are discussed in detail in [14] and [15].

### B. *DT Types and Operation*

There are several types of DT implementations as defined by Grieves and Vickers in [16], each of which serving a specific purpose.

The DT Prototype (DTP) is a digital representation used for collecting data from physical systems or processes, often used as a test before creating other types of DTs. For instance, a manufacturer might use a DTP to collect data from a prototype of a new product to enhance its design and reduce costs. However, it cannot be considered a fully DT as it lacks feedback to the physical system.

The DT Instance (DTI) is a virtual replica of a single physical entity, such as an engine within a complex system. A DTI might be utilized to monitoring the performance of an engine in a specific aircraft, and the data collected from the DTI might be utilized for example to optimize maintenance schedules and improve efficiency.

Extending the DTI concept, the DT Aggregate (DTA), aggregates data from multiple DTI instances, turning into a virtual model of the physical system or process, rather than on a specific instance of it. For instance, a manufacturer might use a DTA to monitor the performance of hundreds of engines located in different places, to identify recurrent issues and optimize maintenance schedules.

From a distinct perspective, recognized the fact that the environment can notably impact the DT performance, the term DT Environment (DTE) refers to the virtual replica of the conditions a physical system operates in. Thus, a DTE might be utilized to simulate the effects of using a DTI in different locations and test different weather conditions to assess the performance of the beforementioned engine, as it might not be the same landing in a northern country or in a tropical one.

The DT operation is illustrated in Figure 2, wherein sensory data collected from various sources within the physical entity (depicted in light grey) such as speed, temperature, and pressure readings, as well as data about the environment, can be transmitted to the DT (represented in dark grey) via a physical-to-virtual communication channel (detailed in Figure 3). These data enable simulations and analytics to be performed within the virtual instance. For example, a DTI and DTP may be utilized to assess various configurations of a manufacturing process in order to determine the optimal setup. In a DTI, interventions can be initiated based on DT feedback using the virtual-to-physical communication channel (detailed in Figure 3) and subsequently implemented in the physical system if necessary. For instance, whenever a process modifies any aspect of the physical or virtual system, this information is exchanged between them, enabling the necessary actions to be taken. If a malfunction is detected in the physical system, the DTI can be utilized to identify the root cause and propose potential remedial actions, which can then be implemented in the physical system

The design phase of a DT is a crucial challenge, as it requires a comprehensive understanding of the system to be represented and the selection of the relevant parameters to attain the desired level of fidelity. Fidelity refers to the level of detail and precision with which the DT represents the physical system or process, as well as the number and type of parameters that are transferred between the digital and physical environments. Last but not least, the performance should also be considered, referred to as the bandwidth needed or the required computing time. Indeed, this is extremely important as DT implementations usually are highly demanding in resources. For example, if the goal of the DT is to optimize the maintenance schedule for a specific type of engine, it would be important to include a wide range of parameters related to the engine's performance and operating conditions in order to achieve a high level of fidelity. On the other hand, if the goal is simply to monitor the overall performance of the engine as a part of a more complex vehicle DT, a lower level of fidelity may be sufficient to obtain a better performance.

In order to ensure that the DT can accurately represent the physical system, and, consequently, meet the goals of the design, it is important to have highly skilled specialists engaged into the design phase. For instance, a team of engineers with expertise in the design and operation of a specific engine type, working together with computer scientists and data analysts would be well-suited to contribute to the design of a successful DT for that engine.

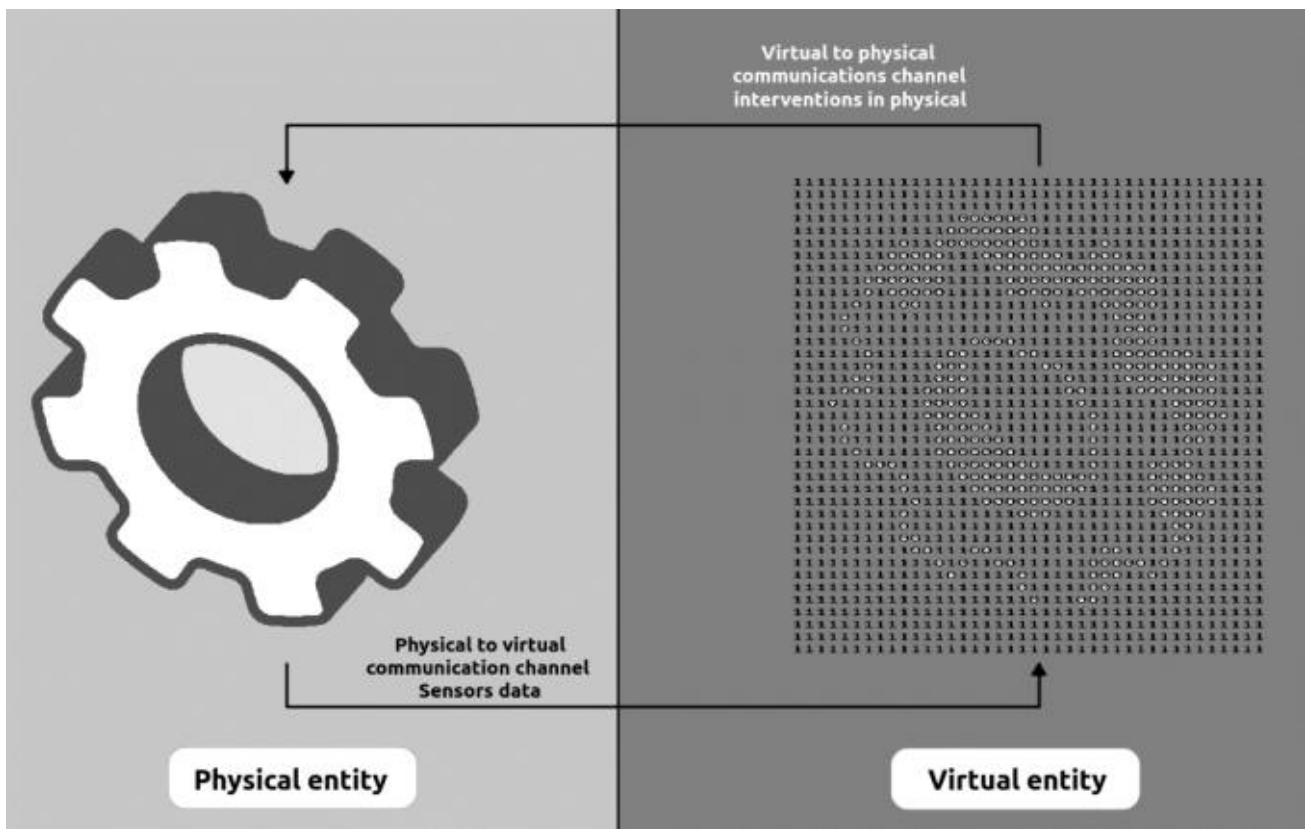

Figure 2 DT basic concept

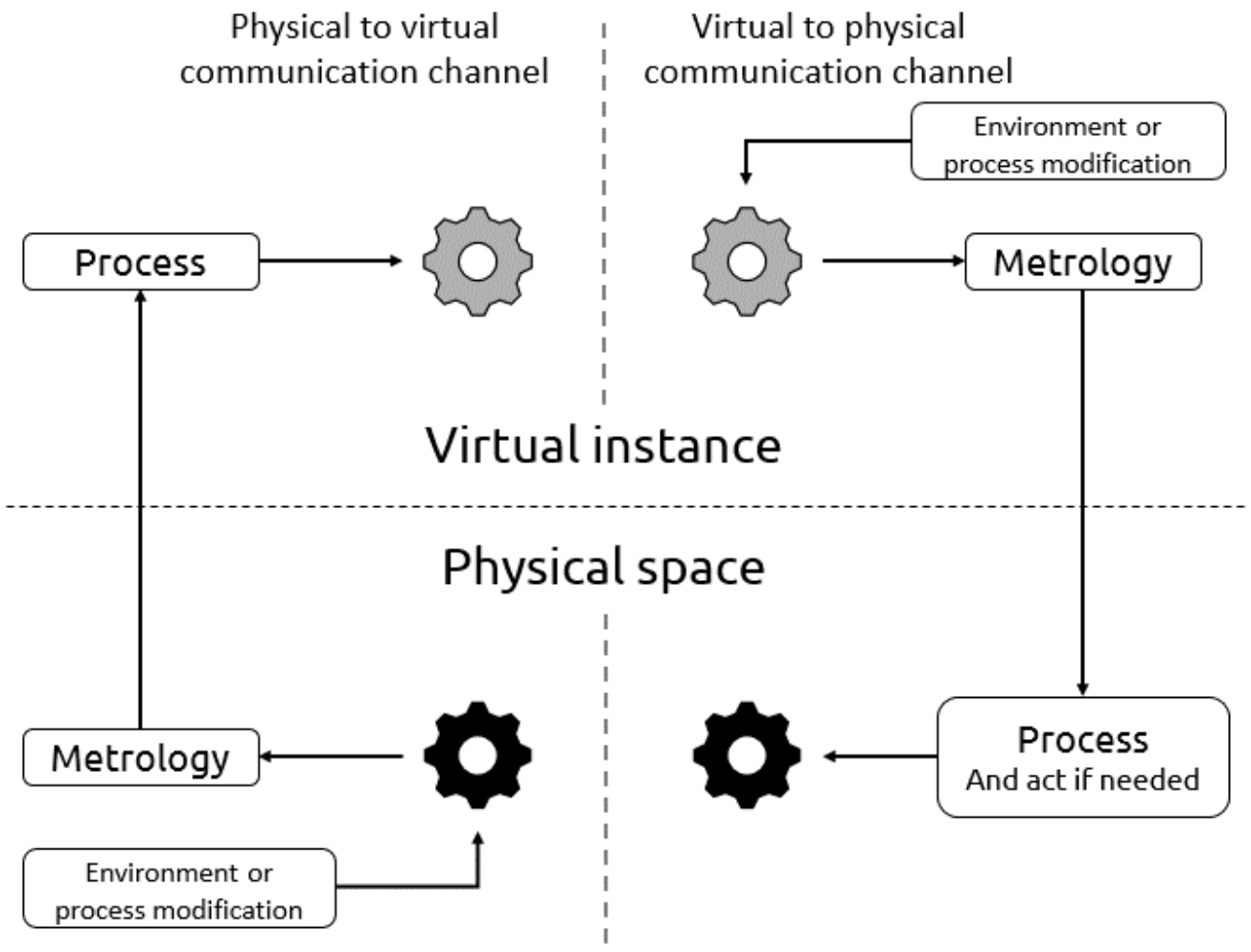

Figure 3 Detail of DT communication channels

The parameters referenced before include types of data, information, and processes. These parameters should be carefully chosen and regularly checked during the operation of the DT as their quantity and relevance can affect the DT's fidelity and performance.

### C. *DT Lifecycle Description*

Figure 4 illustrates the typical procedure for the creation of a DT, which is divided into several phases.

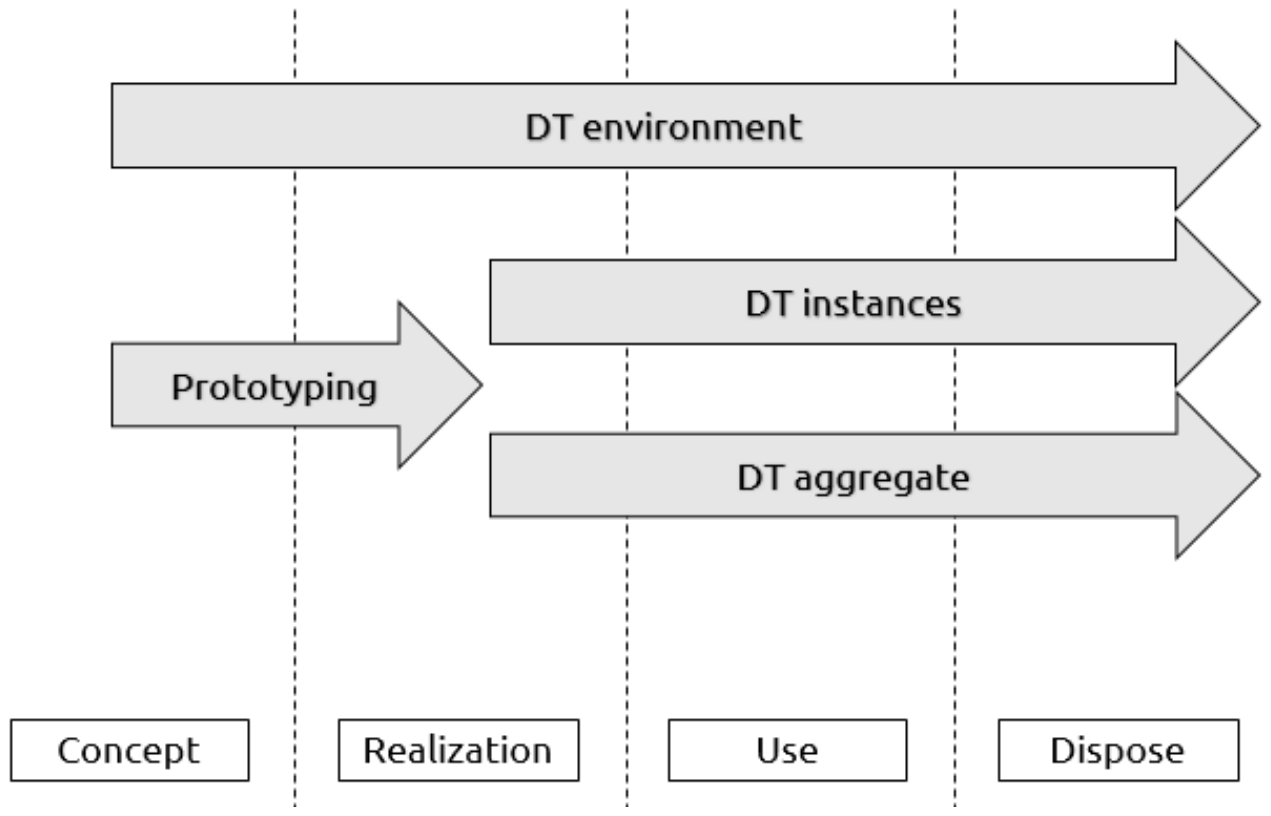

Figure 4 DT lifecycle

During the concept phase, key considerations include the environment where the DT will be deployed in, the virtualization of the physical system or process, and the DTP creation. The parameters collected from the physical environment and its corresponding DT must be carefully evaluated, and in some cases, it may be necessary to add additional sensors to the physical system and its environment in order to improve the DT fidelity. This tuning might not be so easy in some scenarios. For example, while there should be no issues in an Industry 4.0 context (machines are often highly instrumented with many sensors), it may be way difficult in a healthcare context (for example when considering the lack of comfort of some of the wearables that patients are asked to use). The key challenge to address toward a successful DT design is to select the right relevant data for the problem to be modeled, to be properly processed in order to achieve the intended goals, thus avoiding the need for collecting and processing irrelevant data.

During the realization phase, the DTP is developed and extended to create an individual DTI leveraging the data collected from each physical system or process to be replicated. It is also common to deploy a DTA that consumes information from all the DTIs in order to make general improvements to the product or process, detect recurring faults, or even anticipate critical maintenance.

During the support and use phase, all elements within the DT are interconnected, and the information gathered is used to improve both the physical and virtual sides. The DTIs can be refined by evaluating which information from the physical processes is the most relevant, thus reducing the amount of data to be processed and increasing the fidelity of the DT. The physical system or process can also use the information obtained through the digital processes of the DTIs to make necessary adjustments or interventions.

During the retire phase, the DTIs and their corresponding DTA and virtual environment are typically archived, allowing the owner to utilize the valuable information generated throughout the entire DT lifecycle in the future to design similar products or processes, considering the potential issues that the previous version encountered.

### D. *DT Characteristics*

Two main characteristics may be emphasized for a DT to properly and accurately achieve its objectives, i.e., effective communication and proactive performance.

Effective communication is crucial for the success of a DT system. The physical and virtual representations must have a communication system enabling continuous and reliable exchange of information about the processes on each side. While the communication channels may utilize different network architectures and synchronization times, a DT isolated from the real-world data would not be effective. Through the communication channels, the DT receives dynamic information from the physical system or process to keep the virtual instance and its environment updated. Using this data, the DT can provide feedback to the physical system and other physical systems in similar environments. All data received, including data generated by the digital twin's processes, should be stored for the DT to access at any time as historical data.

The other important aspect of the DT is its ability to be updated with new data, thus improving its predictive, corrective, and prescriptive capabilities and accurately reflecting the current state of the physical system or process. To this end, DTs often leverage novel technologies and tools, such as mathematical models, artificial intelligence algorithms, data analysis techniques, data dimensionality reduction, self-adaptation, and self-parameterization, to name a few. By using these technologies, the DT can perform its functions efficiently, achieving real-time cyber-physical synchronization, and users can interact with a virtual representation of the physical system or process to view its state, perform stress tests, or simply modify input parameters to predict the final result.

### E. *Key DT Verticals*

The following subsections present the most common verticals for DT, each introducing some specific use cases.

#### 1) *Manufacturing*

The DT concept has gained widespread adoption in the manufacturing industry [7] mainly because this was the seed vertical other sectors mirrored in. Within the manufacturing sector, DTs are used for two main purposes: products and manufacturing processes. The data collected from DTs help companies to make informed decisions about products and processes, resulting in time and cost savings as well as in increased efficiency.

On the one hand, in the product domain [17], DTs are commonly used to perform simulations [9] with demanding variables (such as speed, resistance, or environment), to achieve an improved performance, a specific safety, or to determine the product limit specifications in different environments. The collected data is also utilized to identify hardware defects and prevent faults through proactive maintenance planning. All data collected from multiple DTIs or their associated DTA is utilized to inform the design of subsequent product versions.

On the other hand, DTs are employed to digitize manufacturing processes, assisting with: i) the management of machinery [18] and personnel workloads; ii) the prevention of occupational risks [19]; iii) the management of space in warehouses and logistics [20], and; iv) the optimization of production machinery speed, performance, and scheduling [21]. In addition to generating valuable information about the manufacturing process, a DT facilitates the rapid resolution of potential problems that may arise, such as the management of available resources following a production line failure. Using the DT technology, multiple alternatives can be computed or simulated to select the new resource assignment that yields the optimal results, mitigating the disruption caused by the failure. DTs also enable the monitoring of the work environment to prevent industrial accidents, alleviate work fatigue, and ensure the safety of human-robot interaction. Many DTs in manufacturing processes utilize artificial intelligence and big data techniques to more efficiently achieve superior results.

The automotive industry [22] is a special case within the manufacturing sector, as product failures can result in significant harm to customers and the repair of minor failures in vehicle components can incur high recall costs for the company. DTs are used to digitally test certain vehicle components (such as engines, shocks, control units) in various environments and with varying requirements to identify design, hardware, or recurring faults prior to the finalization of a model. Once the product is complete, DTIs can also be created with their corresponding DTA to collect data for the detection of recurring failure patterns or the improvement of future designs.

Aviation is another use case in the manufacturing industry [23], extending its relevance even further due to the paramount importance of safety. Indeed, DTs can reduce design time in the aircraft design stage, as aircraft structures and systems are complex and must be highly reliable. DTs can simulate the behavior of aircraft structures under a variety of conditions, the response of safety systems to failures, and the execution of a complex set of tests. These tests would be time-consuming and costly to perform using alternative methods. During the aircraft building process, DTs are utilized to reduce verification costs for certain parts and ensure that the final product meets the requirements of the design phase. Similar to the impact of minor vehicle failures on cost, safety, and brand reliability, these factors are critical in aviation.

Once an aircraft is constructed, accurate fault detection and proactive maintenance planning are both essential due to the high safety standards in the aviation sector [24]. DTIs with their corresponding DTA make it easier for companies to run testing procedures in the digital realm aimed at identifying and preventing malfunctions in aircraft parts or systems, enabling proactive updates to maintenance routines. This information can also be utilized for regular inspections to identify any parts requiring special attention.

Finally, in the aviation field, the DT concept is also utilized for pilot and maintenance personnel training, as well as the improvement of airport management procedures and the enhancement of airport passenger capacity.

#### *2) Healthcare*

DT technology has the potential to revolutionize healthcare in several ways, depending on the element being twinned. DTs in healthcare can be classified into three main categories: patient twinning, facility or service digitization, and trial digitization.

One area of particular interest in healthcare is patient twinning [25], also referred to as personalized or precision medicine [26]. It aims at creating a DT of a patient using its own data. Usually, the DT mirrors a specific organ related to a specific disease, as a full human body DT is hard to develop and extremely consuming in resources. Some ongoing efforts may be found, such as those referring to the heart [27] for cardiac diseases. The DT model can then be used to predict the progression of the disease [28], evaluate the potential impact of a specific treatment on the patient [29], prepare the patient for surgery [30], and test the size of an implant and the optimal procedure for placement according to the tools and sizes available. Some DT models are also generated from data from multiple patients to train medical personnel who are still gaining experience. The successful implementation of these techniques has the potential to optimize healthcare resources, improve accessibility, and enhance the quality of life for patients.

The DT paradigm is also applicable to hospital infrastructure [31] and services, as it is not significantly different from a manufacturing process. Many hospitals have created DTs for individual services or groups of services [32] to improve productivity. Others have created a replica of a patient's behavior during hospital visits to enhance patient care and reduce congestion at patient care desks. By utilizing the DT technology, hospitals have been able to shorten patients' visits by eliminating the need to wait in multiple locations [31].

Finally, in the pharmaceutical industry, DT technology has also brought significant changes. Some companies have used previously collected data to generate models and conduct pre-clinical trials [33], reducing the cost and time of drug development. In other cases, a digital twin based on a patient's data is used to predict which drug will yield the best results for a patient using data previously collected from multiple patients. This approach has the potential to accelerate the patient's treatment process, saving time and avoiding complications from treatments that may not be as effective as the predicted one.

#### *3) Supply Chain*

A supply chain DT is a virtual replica of a company's supply chain, used to analyze and optimize the flow of goods, information, and resources. It allows companies to simulate various scenarios and make informed decisions about their supply chain operations.

There are many benefits in implementing a supply chain DT, summarized as: i) an enhanced visibility and transparency by tracking and analyzing data from the entire supply chain, and; ii) a more comprehensive view of companies’ operations, identifying bottlenecks and inefficiencies, and letting them make better informed decisions about resources’ allocation. Therefore, companies are endowed with higher levels of flexibility and agility, as they can quickly and easily respond to changes in demand, market conditions, and other factors. This can be particularly valuable in times of uncertainty or disruption. For example: i) by identifying and addressing the detected inefficiencies in the supply chain, companies can reduce waste, lower costs, and improve their overall efficiency; ii) simulating various scenarios, companies can identify and mitigate potential risks before they occur, helping to reduce the impact of disruptions and protect their business, and; iii) improving the customer satisfaction by optimizing the supply chain, companies can improve their delivery times and responsiveness, leading to increased customer satisfaction. In short, the use of a supply chain DT can help companies to better understand and optimize their operations, leading to improved efficiency, agility, and competitiveness.

The implementation of a fully digital twinned supply chain [34] is a complex task, but even partial twinning can bring numerous benefits. From a global perspective, the supply chain consists of many retail shops connected to one or more warehouses through various modes of transportation. These warehouses are in turn linked to many other warehouses, some of which are connected to the manufacturers of the products, who require raw materials, etc. Given the diverse types of transportation available, each with its own unique set of advantages and disadvantages, creating a DT of the infrastructure [35] can be highly advantageous in decision-making processes, particularly in the face of issues within the transport chain, production chain, or shifts in demand. Historical data generated by DTs can also aid in the prediction of significant changes and the implementation of mitigating actions in advance.

#### *4) Communication Networks*

The increasing number of devices connected to communication networks, which is expected to grow by billions annually [36], makes the concept of DTs particularly relevant. DTs of network infrastructure can be used in many areas, with a notable focus on improving service quality, detecting anomalies and controlling traffic. First, with the growing number of devices, networks must be constantly adjusted to meet real-time needs, fueling service providers to dynamically increase capacity or redirect traffic as necessary in order to continuously provide the desired quality of service. While this is already being managed to some extent, the use of DT networks (DTNs) allows for the simulation of network upgrades to determine the most beneficial action for the system. DT networks can also be used to simulate an increase in the number of users at a particular point, enabling better planning for network growth. Second, another key application of DT networks is anomaly detection. A properly functioning network should exhibit a behavior similar to that of its DT network, so if both differ, the DT can alert earlier and help to identify the source of the failure. Finally, in terms of data traffic control, a DT network can be beneficial in configuring routes to reduce the load on certain links. Indeed, with real-time traffic information, a DTN can modify routes to reduce the number of hops for transmitted traffic, leading to an increase in network capacity and a decrease in energy costs. Additionally, traffic generated by routing protocols can be handled in the virtual domain and then transferred to physical devices, reducing the load on actual networks.

One major challenge currently facing DT networks is scalability and interoperability, as there is no standardized framework. However, due to the significant benefits of digital replication, there is significant research being conducted to develop DTN with the aim of being successfully deployed in the near future.

#### *5) Grids – Supply Networks*

Obtaining a faithful digital replica of oil and gas networks [37], [38], electric power lines [39] and powerplants [40], water supplies [41] and other types of infrastructure can be highly beneficial and strategic for the management teams. There are many areas in these scenarios a DT may significantly contribute to. Indeed, DTs can be used to monitor and regulate the production or flow of energy or service, as well as identify the most efficient paths for transporting it. In addition, DTs can be used to simulate and address anomalies [42] or leakages [43], in the infrastructure, helping to design well-structured and dimensioned systems that provide the best service at the lowest cost. However, creating a DT for large supply networks can be challenging due to their size and unique characteristics. For example, the implementation for the water supply network of Valencia, Spain [43] required a simplification in order to be usable in real-time due to its DT complexity and the amount of information used.

## III. Material Evaluation in the HDT Field

After briefly introducing and analyzing the applicability of DTs in different verticals, this section focuses on the deployment of human DTs (HDTs), and discusses the results obtained after revisiting existing efforts in many directions and scopes for HDTs. The results have been divided into three groups, the first referring to scientific publications, the second one related to the knowledge transfer (i.e., patents), and the third on industrial initiatives, an area of relevance albeit not much information is disclosed.

### *A. Scientific Publications*

In Table 1 a summary of the papers found is presented. The second column indicates the area in which the HDT is applied, while the third column specifies the objective of the HDT proposal. The fourth column lists the type of data or sensors used to generate the input for the HDT, and the fifth column specifies the type of output generated by the HDT. The last column indicates the technology on which the HDT is based.

Two verticals can be identified in the review results where HDTs are designed and partially developed. The first one is healthcare putting together the highest number of HDT real deployments, reaching out 9 of 15. The second one is the manufacturing vertical, where 6 deployments have been found. Certainly, the surveyed contributions presented in this paper are those we found after an exhaustive monitoring and tracking process, but this does not exclude that a more recent or a not sufficiently disseminated effort might not be reported in this manuscript. Set then by the contributions volume, we start reviewing HDT initiatives in the healthcare area.

Authors in [44] proposed a semi-active DT method for approximating the occlusion of carotid artery stenoses. The DT system uses artificial vision captured through a short, pre-recorded video on a smartphone, rather than receiving real-time data. This is why it is referred to as "semi-active". Once the video is loaded into the system, the system focuses on a region of interest, such as the central forehead or area below the eyes, and defines points to track head movement. After applying digital filtering and principal component analysis (PCA), the resulting vibration is compared to a mechanical model to determine the percentage occlusion. The authors tested the method using both healthy subjects and only one real patient, the data from the latter being replicated to create synthetic patients. The authors concluded that the proposed method shows significant improvement, although further testing is needed to determine its validity and potential usefulness for other disease types.

A framework for utilizing data generated at the edge, such as data collected from monitoring devices connected to a smartphone via Bluetooth or data obtained from phone sensors and social networks, was proposed by the authors in [45]. The goal was to create an artificial intelligence (AI) system that can process these data at the edge, specifically on the smartphone. However, as the project is not yet mature, the authors developed a proof of concept (PoC) based on an inference engine that runs on a smartphone and is trained with public datasets. The PoC was tested on 200 patients, with 166 used for training and 34 for testing, and was able to classify electrocardiogram (ECG) data into myocardial infarction or not with an accuracy of 85.8%, a precision of 95.5%, and a recall of 86.3%.

In [46], the authors aimed to obtain and classify the size of an abdominal aortic aneurysm (AAA) into four categories (healthy, small, medium, and large), using easy-to-access blood pressure waveforms as the main input. To this end, the authors combine several AI techniques. First, they create a patient's DT using patient's profile data, such as height, weight, and mean arterial blood pressure. Using demonstrated empirical relationships, they create a one-dimensional mesh representing the major arteries in the human body. Next, they estimate cardiac flow using a multilayer perceptron. Then, using blood pressure waveforms from the carotid, femoral, and brachial arteries (using only two of these decreases accuracy), they use a recurrent neural network (RNN), specifically a long short-term memory (LSTM) model, to predict blood pressure waveforms at the end of the vessel. Finally, using a convolutional neural network (CNN), the result is classified into one of the aforementioned categories. The accuracy of this system in detecting AAA problems using synthetic data is 99.1%, and its accuracy in determining AAA severity is 97.8%. Two key issues still need to be addressed: developing the system in a real-world environment, and addressing other conditions or diseases that could give similar results in the inverse analysis.

The author in [47] conducted a survey in HDT technology and proposed a structure for constructing HDTs, as well as a set of features that HDTs should possess. The author also presented two projects that were funded by small business innovative research: one on creating patient DTs for scoliosis physiotherapy, and another one on twinning an aircraft pilot. The first project focuses on creating a virtual patient DT using both X-ray images and scans as inputs to help with scoliosis diagnosis and the design and assessment of physiotherapeutic scoliosis-specific exercises. To this end, the author combines all input information to create a skeleton morphing, followed by muscle morphing. The result can be used to help the scoliosis diagnostic and design and assess the physiotherapeutic scoliosis specific exercises prescribed to the patient. The second project, sponsored by the US Air Force, aims at twinning an aircraft pilot using a variety of inputs obtained from wearable sensors and processed using AI. These inputs are applied to the pilot's muscle-skeletal model, anthropometric model, and physiological model to produce outputs such as personalized training to enhance pilot performance, injury mitigation, physiological predictions, and ergonomics for the cockpit and clothing.

Centered on organ replication, authors in [48] presented a dynamic colon model for creating an in-silico test performance model of orally ingested dosage forms inside the gastrointestinal tract. The goal of this model was to facilitate the early stages of testing and fluid modification for specific disease stages. The model variables can be tuned to simulate different conditions in the colon. The model was created using MRI scans, and a Discrete Multiphysics (DMP) simulation was performed to obtain shear rate results for the fluid and its parameters. The simulation is able to replicate the effects of the colon's contractile wall wave propagation speed, media viscosity, and media volume on the mean wall shear rate, similar to results from an in vitro test.

At cellular level, authors in [48] proposed a framework for creating a 3D map of cells from multiplexed images, with the goal of calculating the distance of cell types from endothelial cells and other valuable features. This framework is intended to be a part of a DT for the National Institutes of Health's (NIH) [49] Human Biomolecular Atlas Program (HuBMAP) [50]. Although the input data for this framework is obtained from skin biopsies, it is designed to be applicable to any tissue type. The input data is processed using micro computerized tomography, and a deep learning (DL) encoder-decoder model is used for segmentation and classification. Autofluorescence [51] images were used as a reference to build the digital 3D model. The model was then used to calculate two distances: the distance between the centroid of the nucleus for immune cells, and the distance between the edge of the nearest blood vessel. These results can be compared between different individuals of different ages and states to search for new disease markers. The framework is tested on biopsies from 12 donors ranging in age from 32 to 72, taken from both UV-exposed and non-exposed anatomical regions. While the results of this test were considered relevant and indeed allow for the extraction of many conclusions at the cellular level, they are not discussed in detail in the article. The framework is currently in a development stage and is considered a semi-active DT, as it does not currently provide real-time data. However, once an individualized deployment is released, it has the potential to track the health state of various organs and follow the progression of diseases in each individual using the semi-active paradigm. The main drawback of this proposal is that nowadays, there is no easy way to perform micro CT on tissues without performing a biopsy.

Authors in [52] presented an HDT that combines X-ray navigation and deep learning to create a cyber-human interaction for surgery training. The main focus of the article is on cybersecurity, which is an important aspect of any DT, as virtual entities are vulnerable to various threats. The HDT was developed to simulate surgeries using augmented reality (AR), virtual reality (VR), and mixed reality (MR) modules. The input data for the HDT is obtained from surgery simulation. The authors developed three modules for peg transfer, vessel cutting, and rope handling, and the main inputs for these modules are the total procedure time and the instrument pathway. Each module also uses specific metrics, such as error in clipping and cutting vessels. In the experiments conducted, the results have been positive, with novice doctors obtaining lower scores than expert surgeons before using the HDT training. However, after training with the HDT, the scores of the novice doctors improved, indicating that the software is effective in its goal of improving surgical skills.

On a different health area, a DT framework called SmartFit [53] was designed to support trainers and coaches in optimizing athletes' behavior. SmartFit mirrors the athletes' team using DT techniques and inputs data collected from IoT devices embedded in wearables, as well as data logged from applications tracking athletes' meals and mood. The total number of inputs collected is 22, including data from the applications on meals and mood, and data from IoT sensors on daily steps, number of floors climbed up, walking distance, seating activity, and type and intensity of activity. The output data from SmartFit are rules or suggestions for trainers to pass on to the athletes. The DT analyzes historical data and calculates a score for athletes' inputs, and once sufficient historical data has been collected, the training phase begins. During this phase, the DT's AI component is generated and processes new data to release updated suggestions, which are then stored as historical data. The authors tried the framework using a dataset recorded in 2016 by a teenage male football team over the course of three days, and the system release suggestions to improve scores in areas where the athletes had performed poorly (e.g., increasing minutes of moderate activity on day 3 by 24 units). As future work, the authors plan to record activity over a longer period of time to build a more solid base of historical data and to develop a faster method for generating the DT based on this data.

So far, healthcare related initiatives have been presented. Next, we review HDT initiatives in the manufacturing vertical. As said before, six contributions have been found.

Authors in [54] proposed a system for integrating human workers into a Cyber Physical Production System (CPPS) using HDTs. The goal of the system is to schedule tasks in the CPPS based on the skills, experience, and preferences of the employees. HDTs are used to emulate the behavior of the employees and automatically schedule tasks according to the available machines and the capabilities of the workers. The inputs for the HDTs include the employee's skills, current tasks, experience, and preferred tasks. The system is rule-based, with all necessary information stored in a database that is updated based on the tasks performed by the employees and the time taken to complete them. By using this system, authors aimed to improve the efficiency of the CPPS and the satisfaction of the employees by allowing them to learn new tasks or perform tasks that align with their preferences.

In [55], authors aimed to improve ergonomics and station reorganization in shop floors by using a DT that includes an HDT. To obtain the HDT, the authors collect human motion data using low-cost capture devices, to recognize the tasks being performed (e.g., picking or placing an object, walking, carrying something). The input data consists of human joint coordinates represented in a three-dimensional space, which are then simulated using a data-driven motion synthesis approach. The output of this system is not directly related to the HDT, but rather provides production performance measures, 3D spatial requirements, and a standardized ergonomics score that can be used to improve the ergonomics of the work environment. The proposal was tested in a case study where an operator had to pick and place warehouse components from a rack to a trolley, and the real-world data is collected and transferred to the DT to assess the ergonomics and cycle time for the operation. The scenario is set up in a 10 square meter area with two racks of different heights, and the operation is performed by three actors, each performing the task five times. The transformation of the real input data to the virtual representation takes between 5 and 10 minutes to complete, and multiple simulations need to be run in order to find the configuration that resulted in the best performance. This approach allowed the completion of the design of the workstation in a matter of hours, rather than spending days trying different configurations.

Similarly, the work presented in [56] proposed a framework for improving ergonomics in the workplace. The case study involves an assembly task that places two steel components together and joins them using four screws. The experiment is carried out by an operator wearing a device that collects data on her movements using motion sensors. The data collected by the sensors is sent to a smartphone app, which generated a CSV file containing Euler angles, quaternions, and posture angles for each of the ten segments measured (lower legs, upper legs, forearms, arms, trunk, and pelvis). To simulate the task, the authors used Tecnomatix Process Simulate software by Siemens, incorporating the data collected from the app to generate an accurate simulation. The simulation allows the authors to identify areas in the process where small modifications could improve ergonomics. For example, adding height to a cart reduced the effort required to lift the pieces from a low shelf to within standard ergonomic limits, and replacing a gun screwdriver with an angle screwdriver reduced the counter reaction force to within acceptable limits. This case study demonstrates the usefulness of the system in accurately assessing the ergonomics of a workplace and in designing work stations, particularly when combined with a DT of the workplace.

Aligned to the references above, authors in [57] proposed a method for designing and evaluating human-robot collaboration tasks in manufacturing environments using a DT. The goal is to improve efficiency and ergonomics by allowing the robot to handle the heavier tasks and the human to handle tasks the robot is unable to perform. In the event that the robot encounters any issues with a task, it can skip the task and delegate it to the human, rescheduling the plans for both accordingly. The DT system allows for real-time integration of the work environment, the robot, and the human, and is used to guide the task rescheduling process. To create the HDT, the authors used a Japanese body dimension database to estimate the body dimensions based on the worker's height and weight, and generate a link structure using the body joints. A digital deformable skin surface is then applied to the link structure. Input data for the HDT is collected using a marker-based optical motion capture system called OptiTrack, in combination with multiple cameras placed in the work area. The HDT is integrated into the shop floor DT space and serves as a guide for the task rescheduling system. The system is tested in a work environment designed for picking parts, with shelves on either side containing 40 boxes of parts. A total of 37 reflective markers are attached to the worker's clothing and tracked by 16 cameras mounted on the ceiling. A magnetic tape is also placed on the floor to guide the robot. An initial schedule is provided to the system, which adapts dynamically based on the human cycle time, any issues the robot encounters with picking parts, and HDT ergonomic assessments. The system can display the DT on a screen, with the human parts colored according to joint effort.

Different to the references before, authors in [58] presented a worker DT focused on employee preferences. The main goal is to schedule factory tasks to the most suitable employees, considering factors such as their skills, preferences, character, motivation, and mood. The inputs for the HDT implementation are collected manually and stored in a database, and some of them are updated based on the employee's machinery use hours. The HDT is rule-based and will output an acceptance or rejection of a task based on the input and the HDT's properties. In a lab test, the authors replicate a factory 4.0 workplace with an employee picking parts from a flow station and a shelf to join the parts. When a new task is added to the factory flow, the system first checks the availability of materials to perform the task. If the materials are available, the system asks the HDTs of employees who have the skills or are willing to learn new operations if they can fit the new task into their schedule. With all affirmative answers from the HDTs, the system determines the best way to share the tasks between employees, such as through parallel production. If the system only receives rejections from all HDTs, it sends a broadcast to all physical employees to ask for the new task. This HDT implementation may not be very complex, but it demonstrates how employee wellbeing in the workplace can be improved in a factory by allowing employees to choose their preferences, learn new tasks, and have tasks distributed in the most effective way. This can lead to increased productivity as employees can do tasks they are familiar with, learn new ones, or alternate between them, and can also improve ergonomics by allowing for the easy switching of tasks to prevent repetitive movements.

In [59], a system for real-time calculation of safety distance in human-robot collaboration was presented. The system captures the shared workspace between the human and the robot using RGBD sensors. One sensor is used to scan a section of the physical working environment, while the second sensor scans the other section and tracks the human's skeletal information. The RGBD images are sent to a deep learning server, which uses machine learning algorithms to generate a point cloud for the robot and a 3D simulation of the human, including the skeletal data. The simulation is then used to calculate the safety distance between the human and the robot. The output of the system is provided to the robot, which can be instructed to stop or slow down based on the risk, and to the human, who can be warned through MR glasses. The system aims to improve safety and efficiency in human-robot collaboration by enabling the robot to adjust its behavior based on the real-time position and movement of the human in the shared workspace.

Table 1. Summary of papers in the HDT field

| **Ref** | **Area** | **Goal** | **Input** | **Output** | **How it's processed** |
|---|---|---|---|---|---|
| [44] | H | Approximate severity of carotid artery stenoses | Patient anthropometric data and face video | Carotid occlusion percentage | Artificial vision and computational mechanics |
| [45] | H | Ischemic Heart Disease (IHD) detection | ECG | Classification of data in Myocardial Infarction or not | AI 2 CNNs pipeline |
| [46] | H | Abdominal aortic aneurysm detection and classification | RNN + CNN | AAA diameter | AI RNN + CNN |
| [47] | H | Design and prescribe Physiotherapy Scoliosis Specific Exercises | Body scan, X-ray, skeletal templates | 3D model | Simulations |
| [47] | H | Obtain a military aircraft pilot replica | Many body models and sensors | Performance degradation and impeding casualty | Artificial intelligence |
| [60] | H | Replicate colon dynamics | Magnetic resonance imaging, fluids variables | Colon absorptions | Discrete Multiphysics simulation |
| [48] | H | distance distributions for cell in three dimensions for 3D digital skin biopsy data | images of cells taken from biopsies | interactive visualization in 3d of the cells | Extracting high-quality images |
| [52] | H | Predict health anomalies | Labeled clinical data | A prediction model | CNN |

| Ref | Area | Goal | Input | Output | How it's processed |
|---|---|---|---|---|---|
| [53] | H | Increase athlete's performance | Food income, activity sleep. gathered from IoT and manual way | Suggestions to optimize athlete's behavior | AI using historic data |
| [54] | M | Schedule tasks in a factory depending on HDT | Worker preferences | Task acceptation or rejection | Rule based |
| [55] | M | Improve ergonomics and station reconfigurations | IMU, optical and tool sensors | Motion simulations | Simulations |
| [56] | M | Process ergonomics assessment | IMU | Ergonomic indexes | Simulations |
| [57] | M | Production efficiency and ergonomics | Video, MoCap, IMU | Task schedule, ergonomics assessment | Rule based schedule and DH software DhaibaWorks |
| [58] | M | Task scheduling in factories according to expertise and openness | Worker skill, preferences, character, expertise, mood, motivation | Task schedule | Rule based decision making |
| [59] | M | Safe human robot interaction | RGBD sensors | Safety distance for the robot to stop or reduce speed | Artificial vision, deep learning and 3d point cloud |

*H healthcare, M manufacturing*

## B. *Intelectual property: Patents*

This section presents the included results from patent analysis exposed in section 1d. Most of the patents are authored by institutions from China and Korea. The patents that pass the screening phase followed in this survey are listed in Table 2, along with their respective focus area and objective. A word cloud graph in Figure 5 illustrates the most frequently occurring words in the titles, having a bigger size for the most repeated words.

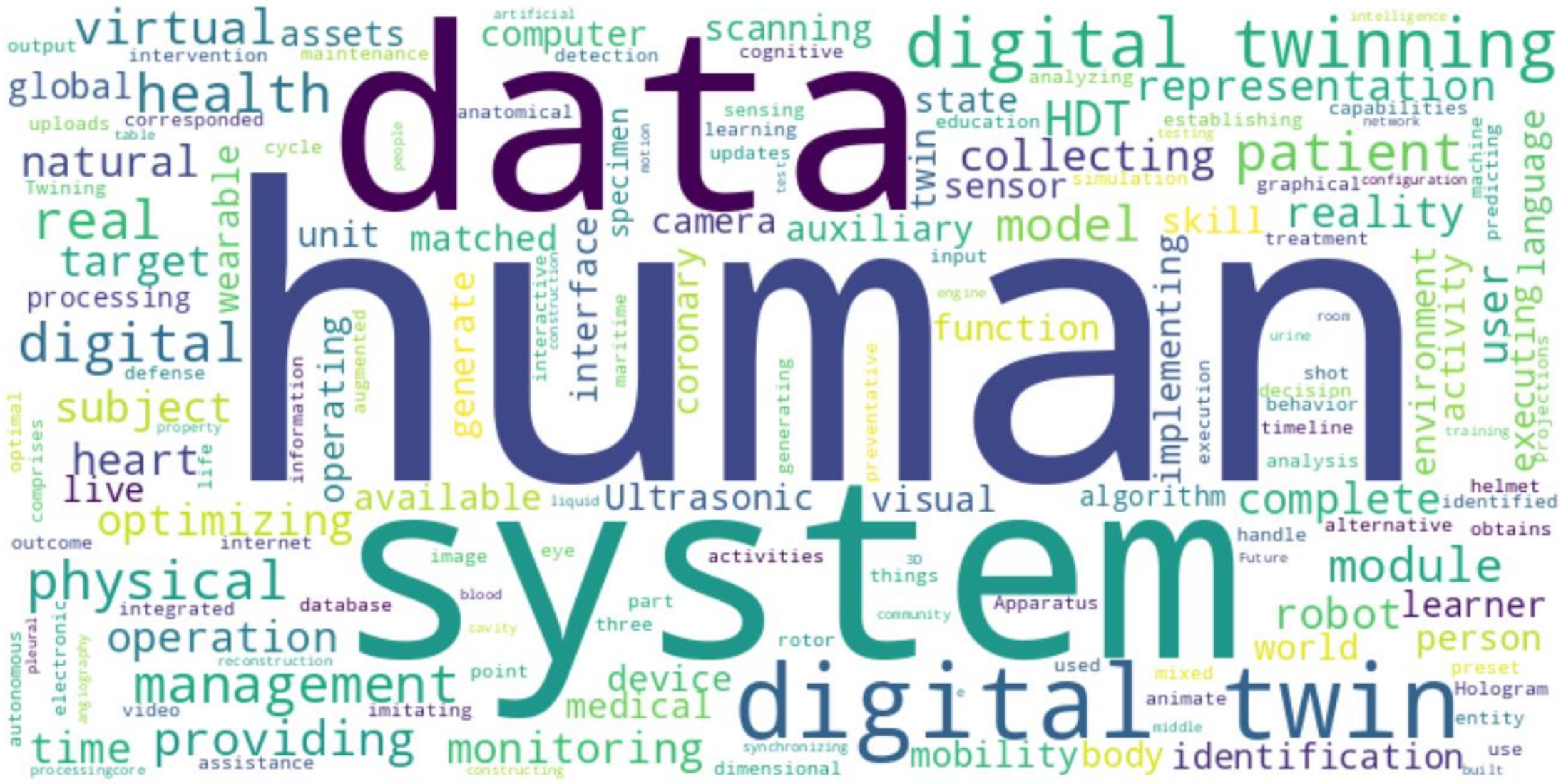

Figure 5. Patent titles word cloud

Table 2. Summary of patents in the DT field

| Reference | Area |
|---|---|
| [61] | Shape selection human eyes |
| [62] | Human Robot Interaction |
| [63] | Healthcare |
| [64] | Human robot interaction |
| [65] | Military |
| [66] | Healthcare |
| [67] | Human activities |
| [68] | Healthcare |

| Reference | Area |
|---|---|
| [69] | Healthcare |
| [70] | Maritime transportation |
| [71] | Learning HDT |
| [72] | Healthcare cardiology |
| [73] | Human rescue |
| [74] | Healthcare |
| [75] | Healthcare |

### C. *Industrial Initiatives*

The most relevant projects undertaken by the industry sector found in the technology radar process building this survey are discussed in this section. As many of these projects have limited scientific literature available, the websites of the products are referenced, along with a brief overview of the details obtained from each project.

Semic Health Digital Body Total [76] is a DT of human biological systems, organs, or molecular systems that utilizes artificial intelligence (AI) to aid in the diagnosis of medical conditions and predict possible health complications. It is an evolution of the Cyber Bio Twin project [77] , which aimed to create a digital replica of human biological systems for the purpose of improving healthcare. The Digital Body Total software is based on the collection of data from various subjects to create computational and mathematical models that replicate the physiology of human biological systems. These HDT models are created with the aid of AI. Digital Body Total has obtained approval for use in the US healthcare system by the US Food and Drug Administration (US FDA). This approval highlights the potential value as a tool for healthcare professionals to improve the accuracy of diagnoses and predictions, as well as its adherence to regulatory standards. Digital Body Total software is a valuable tool that can aid in the analysis and management of clinical trials, disease management, outbreak prevention and management, and lifestyle management, while ensuring the privacy and security of an individual's personal medical data. Moreover, the HDT can be modified to examine and predict the effect of changes on an organ and can be experimented on by introducing foreign substances to test their effects. It can also be modeled to consider environmental and medical factors such as air composition and medication or nutrient intake. This allows healthcare professionals to better understand the functions and dysfunctions of various organs and systems within the body and develop more effective treatments and therapies.

Sim&Cure Sim&Size [78] software is used to generate three-dimensional (3D) models of the patient's brain and blood vessels, which allows surgeons to visualize the aneurysm and plan the surgical approach. The software may also include simulation capabilities, enabling surgeons to practice the surgery virtually and test different options before performing the procedure. The Sim&Size software goal is to improve the accuracy and success rate of brain aneurysm surgeries by providing interactive visualization and simulation tools that allow surgeons to more effectively plan and execute the procedure.

Focusing on the heart arena, Dassault Systemes presented the Simulia Living Heart [79], a high-fidelity heart model that converts 2D scans into a 3D model. The Living Heart project has signed a 5-year research agreement with the FDA, followed by a 5-year extension, with the goal of developing and validating highly accurate personalized digital human heart models. These models will serve as a common technology base for education and training, medical device design, testing, clinical diagnosis, and regulatory science, with the aim of rapidly translating current and future innovations into improved patient care. For example, the use of a 3D heart model to perform in silico tests can reduce the need for animal testing and the number of patients subjected to potentially harmful early stage trials, thus saving a significant amount of time and resources in the clinical trial process for medical innovations.

Also dealing with heart, Siemens Healthineers [80] presented in the Radiological Society of North America (RSNA) conference in 2018 its technology [81] based on the use of Magnetic Resonance imaging and ECG to simulate heart physiological processes through a DT. This approach was used at the University of Heidelberg to predict the outcomes of various treatments and select the one that best fits the patient. Cardiologists have also utilized this method in the context of cardiac resynchronization therapy, a treatment for chronic congestive heart failure, in which electrodes are implanted on the heart right and left ventricles to resynchronize beating. By training a heart DT with a large amount of data and virtually implanting electrodes to deliver virtually generated pulses to the heart, the potential success of the therapy can be predicted. If the asynchronous pumping of the virtual heart is corrected, it serves as an indication that resynchronization therapy could also be successful in the real patient.

Last initiative also in the heart arena, seats on Philips developments, namely Phillips Heart Navigator [82] and Phillips Heart Model [83], that are medical software tools that utilize a generic heart model combined with patient-specific information, such as computed tomography (CT) scans, live X-ray images, and ultrasound data, to generate a three-dimensional (3D) model of the patient's heart. This model is used to assist surgeons in planning and performing transcatheter aortic replacements (TAR), a type of minimally invasive heart surgery, by helping to select the most appropriate device for the patient and the optimal insertion angle. The Heart Navigator also provides live image guidance during TAR procedures to support the precise positioning of the device. The ultimate goal of these tools is to improve the accuracy and success rate of TAR procedures. In the future, Phillips plans to develop augmented reality (AR) models for practicing TAR procedures and for use in the operating room. These AR models will be patient-specific, allowing for more personalized and accurate guidance during surgery.

## IV. Conclusions

After the thorough review carried out in this survey, we may with no doubt conclude that HDTs are a growing area of research with a strong potential to revolutionize the way we approach healthcare, manufacturing, transportation, and many other fields, by simulating and analyzing real-world human bodies and behaviors, aimed at improving decision-making, optimize processes, and enhance safety and efficiency. We may also emphasize that even though HDT applicability ranges a wide spectrum of verticals, healthcare comes up as a highly promising area, being the area where HDT initiatives are deployed the most. Also, there is no doubt about the positive impact that HDT can have in healthcare and in the society as a whole. Indeed, the healthcare sector shows 9 HDT deployments as found in this survey, with a broader scope, ranging from estimating the occlusion of carotid artery stenoses to track football team players' performance.

However, although some deployments are even at a commercial stage, there are some aspects that remain unsolved and thus, still require some attention from the scientific community. Indeed, a lack of contributions in these aspects is strongly hampering the chances for HDT to be widely adopted.

One of the major challenges technology progress must face today, is the need for robust cybersecurity and data privacy protection in the face of the constant, non-stoppable and ever-growing cyber threats. Hackers and cybercriminals are continually seeking out new ways to access sensitive information and steal data, which can have serious consequences for both individuals and organizations. An HDT, considered as a virtual instance, inherently poses many concerns about data privacy and security, as the data used to create the virtual representation of humans must be properly protected to prevent unauthorized access or a non-desired misuse. A parallel issue refers to the lack of transparency around how companies use and share personal data, which can leave individuals feeling uncertain about their privacy online. The use of connected devices, also known as the Internet of Health (IoH), adds an additional layer of complexity to cybersecurity concerns, as these devices are often equipped with weak security and can be exploited by hackers.

In order to address these challenges, it is necessary to prioritize cybersecurity in HDT systems design and development. To this end, encryption techniques and secure communication protocols to send and share data may be used along with prioritizing data processing at the edge rather than relying on centralized servers. Data privacy must also be considered during the development, through the use of "data privacy by design" approaches. Ethical considerations are also crucial in the developments, as they raise important questions about privacy violations, discrimination, and other issues. Researchers and developers must engage with ethicists and other experts in these fields and seek input from stakeholders to ensure that HDT technology is used in a responsible and beneficial manner.

Another important aspect to be considered refers to legal aspects. One of the main concerns is the issue of data privacy and protection as commented before. Different laws in many countries regulate the collection, use, and storage of personal data. These laws often require companies and organizations to obtain consent from individuals before collecting their data and to protect it from unauthorized access or misuse. This is critically related to identifying who owns the HDT. Indeed, many concerns may pop up on who owns the data used to create a HDT and who has the right to use or control the twin. This could be an issue if a company creates a HDT or a DT using data collected from an individual without their knowledge or consent. There may also be legal implications related to the use of HDTs in decision-making processes or other situations where they could potentially affect individuals' rights or interests. For example, if a company uses a HDTs to make decisions about hiring or promotion, there may be concerns about discrimination or bias.

Browsing the current technology progress, we notice that several avenues for future research in the field of HDTs are open. One area would focus on the improvement of HDT accuracy and realism in order to better mimic the behavior and characteristics of real human systems. Research should focus on enhancing current state-of-the-art in HDT modeling and simulation techniques, aimed at creating more useful and relevant HDTs, always considering the main final goal, that is have a fully twinned human able to be used for many purposes.

Another research area would be the expansion of HDT applications to new fields and sectors. Although HDTs have already been applied mainly in healthcare and manufacturing sectors, many other areas may benefit from HDTs. Hence, research is critical to identify the areas where HDT may be beneficial the most. That said, it is also important to figure out which areas would also bring a commercial and large societal impact, indeed, some work have started in clothes design or medical devices personalization. As an example, in a society where several companies and government entities employ the use of DTs, these digital representations could interact directly with a user DT. For this interaction to be effective, the user's digital twin must be able to accurately represent the user's needs and habitual decisions, and be able to make autonomous decisions and carry out tasks that improve the user's overall quality of life. The application of HDT technology can be utilized in situations where there is a human-digital interaction. However, before reaching this goal, the technology must continue to mature and overcome its current limitations, through the completion of ongoing research projects that aim to advance the state-of-the art.

Taking a non-technological perspective, a key area where efforts are required is related to the management of ethical and legal concerns, to ensure that HDTs are used in a responsible and beneficial manner. Particular work in this area should go on generating frameworks or roadmaps to help the HDT technology to be more secure, regulatory compliant, to respect individuals' privacy rights, and to minimize the potential for bias and discrimination.

Finally, the major technical challenge HDT must face refers to achieving much better efficiency and scalability. This is especially important given the resource-intensive nature of HDTs developments, which often require large amounts of data and intensive computing power. To address these challenges, there is a need for more efficient HDT implementations, scalable systems that can effectively manage vast amounts of information in networks and process them in an efficient way to give real time results.

As a summary, any effort towards facing these challenges, will help to benefit from the full potential of HDT systems, supported by a secure, ethical, efficient and accurate adoption of HDTs in many verticals.

## Acknowledgment

This work has been supported by the Spanish Ministry of Science and Innovation under contract PID2021-124463OB-I00, the Catalan Government under contract 2021 SGR 00326 and the Catalan Department of Research and Universities.

## References

[1] B. R. Barricelli, E. Casiraghi, and D. Fogli, "A survey on digital twin: Definitions, characteristics, applications, and design implications," *IEEE Access*, vol. 7. Institute of Electrical and Electronics Engineers Inc., 2019. doi: 10.1109/ACCESS.2019.2953499.

[2] M. Singh, E. Fuenmayor, E. P. Hinchy, Y. Qiao, N. Murray, and D. Devine, "Digital Twin: Origin to Future," *Applied System Innovation 2021, Vol. 4, Page 36*, vol. 4, no. 2, p. 36, May 2021, doi: 10.3390/ASI4020036.

[3] M. Liu, S. Fang, H. Dong, and C. Xu, "Review of digital twin about concepts, technologies, and industrial applications," *J Manuf Syst*, vol. 58, pp. 346–361, Jan. 2021, doi: 10.1016/J.JMSY.2020.06.017.

[4] D. Jones, C. Snider, A. Nassehi, J. Yon, and B. Hicks, "Characterising the Digital Twin: A systematic literature review," *CIRP J Manuf Sci Technol*, vol. 29, pp. 36–52, May 2020, doi: 10.1016/J.CIRPJ.2020.02.002.

[5] C. Semeraro, M. Lezoche, H. Panetto, and M. Dassisti, "Digital twin paradigm: A systematic literature review," *Comput Ind*, vol. 130, Sep. 2021, doi: 10.1016/J.COMPIND.2021.103469.

[6] Y. Lin *et al.*, "Human Digital Twin: A Survey," Dec. 2022, doi: 10.48550/arxiv.2212.05937.

[7] W. Shengli, "Is Human Digital Twin possible?," *Computer Methods and Programs in Biomedicine Update*, vol. 1, p. 100014, Jan. 2021, doi: 10.1016/J.CMPBUP.2021.100014.

[8] M. E. Miller and E. Spatz, "A unified view of a human digital twin," *Human-Intelligent Systems Integration 2022 4:1*, vol. 4, no. 1, pp. 23–33, Mar. 2022, doi: 10.1007/S42454-022-00041-X.

[9] Michael Grieves, "Digital Twin: Manufacturing Excellence through Virtual Factory Replication," *Whitepaper*, 2003.

[10] M. Shafto, M. C. Rich, D. E. Glaessgen, C. Kemp, J. Lemoigne, and L. Wang, "DRAFT MoDeling, SiMulATion, inFoRMATion Technology & PRoceSSing RoADMAP Technology Area 11," 2010.

[11] "DIGITAL TWIN: DEFINITION & VALUE An AIAA and AIA Position Paper," 2020.

[12] R. Stark and T. Damerau, "Digital Twin," in *CIRP Encyclopedia of Production Engineering*, Berlin, Heidelberg: Springer Berlin Heidelberg, 2019, pp. 1–8. doi: 10.1007/978-3-642-35950-7_16870-1.

[13] C. K. Lo, C. H. Chen, and R. Y. Zhong, "A review of digital twin in product design and development," *Advanced Engineering Informatics*, vol. 48, p. 101297, Apr. 2021, doi: 10.1016/J.AEI.2021.101297.

[14] J. Trauer, S. Schweigert-Recksiek, C. Engel, K. Spreitzer, and M. Zimmermann, "WHAT IS A DIGITAL TWIN? - DEFINITIONS and INSIGHTS from AN INDUSTRIAL CASE STUDY in TECHNICAL PRODUCT DEVELOPMENT," in *Proceedings of the Design Society: DESIGN Conference*, 2020, vol. 1, pp. 757–766. doi: 10.1017/dsd.2020.15.

[15] W. Kritzinger, M. Karner, G. Traar, J. Henjes, and W. Sihn, "Digital Twin in manufacturing: A categorical literature review and classification," in *IFAC-PapersOnLine*, Jan. 2018, vol. 51, no. 11, pp. 1016–1022. doi: 10.1016/j.ifacol.2018.08.474.

[16] J. Vickers and M. Grieves, "Digital Twin: Mitigating Unpredictable, Undesirable Emergent Behavior in Complex Systems," 2017, doi: 10.1007/978-3-319-38756-7_4.

[17] C. K. Lo, C. H. Chen, and R. Y. Zhong, "A review of digital twin in product design and development," 2021, doi: 10.1016/j.aei.2021.101297.

[18] G. S. Martínez, S. Sierla, T. Karhela, and V. Vyatkin, "Automatic generation of a simulation-based digital twin of an industrial process plant," *Proceedings: IECON 2018 - 44th Annual Conference of the IEEE Industrial Electronics Society*, pp. 3084–3089, Dec. 2018, doi: 10.1109/IECON.2018.8591464.

[19] G. P. Agnusdei, V. Elia, and M. G. Gnoni, "A classification proposal of digital twin applications in the safety domain," *Comput Ind Eng*, vol. 154, Apr. 2021, doi: 10.1016/J.CIE.2021.107137.

[20] S. Yevgenievich Barykin, "THE PLACE AND ROLE OF DIGITAL TWIN IN SUPPLY CHAIN MANAGEMENT," *Marketing Management and Strategic Planning*, vol. 20, no. 2, 2021.

[21] Y. Fang, C. Peng, P. Lou, Z. Zhou, J. Hu, and J. Yan, "Digital-Twin-Based Job Shop Scheduling Toward Smart Manufacturing," *IEEE Trans Industr Inform*, vol. 15, no. 12, pp. 6425–6435, Dec. 2019, doi: 10.1109/TII.2019.2938572.

[22] D. Piromalis and A. Kantaros, "Digital Twins in the Automotive Industry: The Road toward Physical-Digital Convergence," *Applied System Innovation 2022, Vol. 5, Page 65*, vol. 5, no. 4, p. 65, Jul. 2022, doi: 10.3390/ASI5040065.

[23] H. Aydemir, U. Zengin, U. Durak, and S. Hartmann, "The digital twin paradigm for aircraft – review and outlook," *AIAA Scitech 2020 Forum*, vol. 1 PartF, pp. 1–12, 2020, doi: 10.2514/6.2020-0553.

[24] C. Li, S. MahaDeVan, Y. Ling, S. Choze, and L. Wang, "Dynamic Bayesian Network for Aircraft Wing Health Monitoring Digital Twin," *https://doi.org/10.2514/1.J055201*, vol. 55, no. 3, pp. 930–941, Jan. 2017, doi: 10.2514/1.J055201.

[25] W. Shengli, "Is Human Digital Twin possible?," *Computer Methods and Programs in Biomedicine Update*, vol. 1, p. 100014, Jan. 2021, doi: 10.1016/J.CMPBUP.2021.100014.

[26] K. P. Venkatesh, M. M. Raza, and J. C. Kvedar, "Health digital twins as tools for precision medicine: Considerations for computation, implementation, and regulation," *npj Digital Medicine 2022 5:1*, vol. 5, no. 1, pp. 1–2, Sep. 2022, doi: 10.1038/s41746-022-00694-7.

[27] J. Corral-Acero *et al.*, "The 'Digital Twin' to enable the vision of precision cardiology," *Eur Heart J*, vol. 41, no. 48, pp. 4556–4564, Dec. 2020, doi: 10.1093/EURHEARTJ/EHAA159.

[28] I. Voigt, H. Inojosa, A. Dillenseger, R. Haase, K. Akgün, and T. Ziemssen, "Digital Twins for Multiple Sclerosis," *Front Immunol*, vol. 12, p. 1556, May 2021, doi: 10.3389/FIMMU.2021.669811/BIBTEX.

[29] B. Björnsson *et al.*, “Digital twins to personalize medicine,” *Genome Med*, vol. 12, no. 1, pp. 1–4, Dec. 2019, doi: 10.1186/S13073-019-0701-3/FIGURES/1.
[30] H. Garg, “Digital twin technology: Revolutionaryto improve personalized healthcare,” *Science Progress and Research (SPR)*, vol. 1, no. 1, pp. 32–34, Apr. 2021, doi: 10.52152/SPR/2021.105.
[31] A. Karakra, F. Fontanili, E. Lamine, and J. Lamothe, “HospiT’Win: A predictive simulation-based digital twin for patients pathways in hospital,” *2019 IEEE EMBS International Conference on Biomedical and Health Informatics, BHI 2019 - Proceedings*, May 2019, doi: 10.1109/BHI.2019.8834534.
[32] Y. Peng, M. Zhang, F. Yu, J. Xu, and S. Gao, “Digital Twin Hospital Buildings: An Exemplary Case Study through Continuous Lifecycle Integration,” *Advances in Civil Engineering*, vol. 2020, 2020, doi: 10.1155/2020/8846667.
[33] R. M. C. Portela *et al.*, “When Is an In Silico Representation a Digital Twin? A Biopharmaceutical Industry Approach to the Digital Twin Concept,” *Adv Biochem Eng Biotechnol*, vol. 176, pp. 35–55, 2021, doi: 10.1007/10_2020_138.
[34] J. A. Marmolejo-Saucedo, M. Hurtado-Hernandez, and R. Suarez-Valdes, “Digital Twins in Supply Chain Management: A Brief Literature Review,” *Advances in Intelligent Systems and Computing*, vol. 1072, pp. 653–661, 2020, doi: 10.1007/978-3-030-33585-4_63/COVER.
[35] Y. Wang, X. Wang, and A. Liu, “Digital Twin-driven Supply Chain Planning,” *Procedia CIRP*, vol. 93, pp. 198–203, Jan. 2020, doi: 10.1016/J.PROCIR.2020.04.154.
[36] C. Annual and I. Report, “White paper Cisco public,” 2018.
[37] E. LaGrange, “Developing a Digital Twin: The Roadmap for Oil and Gas Optimization,” *Society of Petroleum Engineers - SPE Offshore Europe Conference and Exhibition 2019, OE 2019*, Sep. 2019, doi: 10.2118/195790-MS.
[38] T. R. Wanasinghe *et al.*, “Digital Twin for the Oil and Gas Industry: Overview, Research Trends, Opportunities, and Challenges,” *IEEE Access*, vol. 8, pp. 104175–104197, 2020, doi: 10.1109/ACCESS.2020.2998723.
[39] D. S. Zolin and E. N. Ryzhkova, “Digital Twins for Electric Grids,” *Proceedings - 2020 International Russian Automation Conference, RusAutoCon 2020*, pp. 175–180, Sep. 2020, doi: 10.1109/RUSAUTOCON49822.2020.9208080.
[40] B. Xu *et al.*, “A case study of digital-twin-modelling analysis on power-plant-performance optimizations,” *Clean Energy*, vol. 3, no. 3, pp. 227–234, Nov. 2019, doi: 10.1093/CE/ZKZ025.
[41] P. Conejos Fuertes, F. Martínez Alzamora, M. Hervás Carot, and J. C. Alonso Campos, “Building and exploiting a Digital Twin for the management of drinking water distribution networks,” *https://doi.org/10.1080/1573062X.2020.1771382*, vol. 17, no. 8, pp. 704–713, Sep. 2020, doi: 10.1080/1573062X.2020.1771382.
[42] W. Danilczyk, Y. L. Sun, and H. He, “Smart Grid Anomaly Detection using a Deep Learning Digital Twin,” *2020 52nd North American Power Symposium, NAPS 2020*, Apr. 2021, doi: 10.1109/NAPS50074.2021.9449682.
[43] F. M. Alzamora, M. H. Carot, J. Carles, and A. Campos, “Development and Use of a Digital Twin for the Water Supply and Distribution Network of Valencia (Spain),” 2019, Accessed: Dec. 02, 2022. [Online]. Available: https://www.researchgate.net/publication/340236600
[44] N. K. Chakshu, J. Carson, I. Sazonov, and P. Nithiarasu, “A semi-active human digital twin model for detecting severity of carotid stenoses from head vibration—A coupled computational mechanics and computer vision method,” *Int J Numer Method Biomed Eng*, vol. 35, no. 5, p. e3180, May 2019, doi: 10.1002/CNM.3180.
[45] R. Martinez-Velazquez, R. Gamez, and A. el Saddik, “Cardio Twin: A Digital Twin of the human heart running on the edge,” *Medical Measurements and Applications, MeMeA 2019 - Symposium Proceedings*, Jun. 2019, doi: 10.1109/MEMEA.2019.8802162.
[46] N. K. Chakshu, I. Sazonov, and P. Nithiarasu, “Towards enabling a cardiovascular digital twin for human systemic circulation using inverse analysis,” *Biomech Model Mechanobiol*, vol. 20, no. 2, pp. 449–465, Apr. 2021, doi: 10.1007/S10237-020-01393-6/FIGURES/14.
[47] Z. Cheng, “Human Digital Twin with Applications,” 2022.
[48] S. Ghose *et al.*, “Human Digital Twin: Automated Cell Type Distance Computation and 3D Atlas Construction in Multiplexed Skin Biopsies,” *bioRxiv*, p. 2022.03.30.486438, Mar. 2022, doi: 10.1101/2022.03.30.486438.
[49] “National Institutes of Health (NIH) | Turning Discovery Into Health.” https://www.nih.gov/ (accessed Dec. 11, 2022).
[50] “HuBMAP Consortium – The HuBMAP Human BioMolecular Atlas Program.” https://hubmapconsortium.org/ (accessed Dec. 11, 2022).
[51] M. Monici, “Cell and tissue autofluorescence research and diagnostic applications,” *Biotechnol Annu Rev*, vol. 11, no. SUPPL., pp. 227–256, Jan. 2005, doi: 10.1016/S1387-2656(05)11007-2.
[52] J. Zhang and Y. Tai, “Secure medical digital twin via human-centric interaction and cyber vulnerability resilience,” *https://doi.org/10.1080/09540091.2021.2013443*, vol. 34, no. 1, pp. 895–910, 2021, doi: 10.1080/09540091.2021.2013443.
[53] B. R. Barricelli, E. Casiraghi, J. Gliozzo, A. Petrini, and S. Valtolina, “Human Digital Twin for Fitness Management,” *IEEE Access*, vol. 8, pp. 26637–26664, 2020, doi: 10.1109/ACCESS.2020.2971576.
[54] I. Graessler and A. Poehler, “Integration of a digital twin as human representation in a scheduling procedure of a cyber-physical production system,” *IEEE International Conference on Industrial Engineering and Engineering Management*, vol. 2017-December, pp. 289–293, Feb. 2018, doi: 10.1109/IEEM.2017.8289898.
[55] N. Nikolakis, K. Alexopoulos, E. Xanthakis, and G. Chryssolouris, “The digital twin implementation for linking the virtual representation of human-based production tasks to their physical counterpart in the factory-floor,” *Int J Comput Integr Manuf*, vol. 32, no. 1, pp. 1–12, Jan. 2019, doi: 10.1080/0951192X.2018.1529430.
[56] A. Greco, M. Caterino, M. Fera, and S. Gerbino, “Digital Twin for Monitoring Ergonomics during

Manufacturing Production," *Applied Sciences 2020, Vol. 10, Page 7758*, vol. 10, no. 21, p. 7758, Nov. 2020, doi: 10.3390/APP10217758.

[57] T. Maruyama *et al.*, "Digital Twin-Driven Human Robot Collaboration Using a Digital Human," *Sensors 2021, Vol. 21, Page 8266*, vol. 21, no. 24, p. 8266, Dec. 2021, doi: 10.3390/S21248266.

[58] I. Graessler and A. Poehler, "Intelligent control of an assembly station by integration of a digital twin for employees into the decentralized control system," *Procedia Manuf*, vol. 24, pp. 185–189, 2018, doi: 10.1016/J.PROMFG.2018.06.041.

[59] S. H. Choi *et al.*, "An integrated mixed reality system for safety-aware human-robot collaboration using deep learning and digital twin generation," *Robot Comput Integr Manuf*, vol. 73, Feb. 2022, doi: 10.1016/j.rcim.2021.102258.

[60] M. Schütt *et al.*, "Simulating the Hydrodynamic Conditions of the Human Ascending Colon: A Digital Twin of the Dynamic Colon Model," *Pharmaceutics 2022, Vol. 14, Page 184*, vol. 14, no. 1, p. 184, Jan. 2022, doi: 10.3390/PHARMACEUTICS14010184.

[61] LU R; ZANG J; YANG Y; MENG W and Patent number: CN113687718-A, "Twining camera for computer identification and human eye identification, has digital twinning module performing digital twinning treatment to physical target shot by camera module that obtains corresponded three-dimensional digital target-All Databases," *23 November*, 2021. https://www.webofscience.com/wos/alldb/full-record/DIIDW:202124455E (accessed Dec. 15, 2022).

[62] LU R; ZANG J; YANG Y; MENG W, "Human-machine integrated digital twinning system, has physical robot matched with physical person, virtual end matched with virtual person and virtual robot, and virtual reality device provided with helmet and handle," CN113687718(A), Nov. 2021

[63] WU Y, "Method for establishing digital twinning based on medical health by using electronic device, involves optimizing medical health decision assistance based on digital representation human body sensing cognitive data and digital representation natural entity intervention data," CN114864088-A, Jun. 2022

[64] SUN Z; HUANG Y; XIAO W; LIU H, "Ultrasonic robot autonomous scanning method of imitating human operation, involves obtaining human ultrasonic image by preset scanning skill and human body locating point and analyzing target part by video auxiliary analysis system," CN112151169-A, Dec. 2020

[65] "The genesis project," Feb. 2019.

[66] CHENGPENG Z; GU Q; CHENG M; LI Z; SHEN J, "Computer auxiliary system for preventative maintenance of human health life cycle, has user graphical interface system that performs user input, output and use function, and data collecting system that uploads detection data to system database through internet-of-things," CN114446482-A, May 2022

[67] "Systems and methods for simulation of humans by human twin," Oct. 2019.

[68] "Methods and systems for generating a patient digital twin," Jun. 2017.

[69] NICOLELLA D P; SAYLOR K J; LIBARDONI M J, "Method for predicting activity for animate subject, involves collecting wearable sensor data of subject by using wearable sensor worn by subject during same activity, and generating outcome data of digital twin's behavior during activating," US2019371466-A1, Dec. 2019

[70] SHARMA A, "System for providing complete global maritime mobility, has augmented reality-mixed reality unit for creating live digital twins that are formed as Hologram Projections of available physical assets including humans in real world," IN201811033797-A, Mar. 2020

[71] KIM W T; KIM Y J; JIN K, "System for providing learning content using human digital twin (HDT), has HDT management unit that generates HDT for learner based on learner information and updates HDT based on education environment data," KR2022027449-A, Mar. 2022

[72] LIANG D; MA Z; QIU J; YIN X; WANG L, "Monitoring heart state comprises obtaining real-time heart data of patients, synchronizing pre-built motion state of patient's digital heart and inputting coronary angiography data into coronary 3D reconstruction network," CN111755104(A)CN111755104(B), Oct. 2020

[73] CAI F; MENG W; QIN H; ZHAO Y; ZHOU Z; CHEN X; ZHANG H, "Future community digital twinning management system based on artificial intelligence digital people in property management room, has natural language processing algorithm training middle table for providing natural language processingcore algorithm for skill configuration engine module," CN114548803-A, May 2022

[74] WANG Y, "Digital twinning system for testing specimen, i.e. blood, urine and pleural cavity liquid, has model management module for implementing model construction and model management functions of digital twins in digital twinning system of test specimen," CN114913964-A, Aug. 2022

[75] WANG Y, "Digital twinning method for health big data, involves collecting data, pre-processing data, constructing digital twin, monitoring operation of digital twin and optimizing digital twin," CN114822861-A, Jul. 2022

[76] "SEMIC Health - SEMIC RF - Artificial Intelligence - Strong AI - SaaS." https://www.semic.de/en/ai-io/semic-health (accessed Dec. 19, 2022).

[77] "Human Digital Twin | Cyber Bio Twin." https://www.cyberbiotwin.com/ (accessed Dec. 19, 2022).

[78] "Sim&Size - Sim and Cure." https://sim-and-cure.com/simsize/ (accessed Dec. 19, 2022).

[79] "Living Heart Project | SIMULIA™ - Dassault Systèmes®." https://www.3ds.com/products-services/simulia/solutions/life-sciences-healthcare/the-living-heart-project/ (accessed Dec. 19, 2022).

[80] "A patient twin to protect John's heart." https://www.siemens-healthineers.com/perspectives/patient-twin-johns-heart (accessed Dec. 19, 2022).

[81] Siemens Healthineers, "Background Information", Accessed: Aug. 04, 2022. [Online]. Available: www.siemens-healthineers.com.

[82] "View details of Philips HeartNavigator." https://www.philips.co.uk/healthcare/product/HCOPT20/heartnavigator-planning-and-guidance-software?_ga=2.34332129.1537695488.1659597368-642156031.1659597368 (accessed Dec. 19, 2022).

[83] "How a virtual heart could save your real one - Blog | Philips." https://www.philips.com/a-

w/about/news/archive/blogs/innovation-matters/20181112-how-a-virtual-heart-could-save-your-real-one.html (accessed Dec. 19, 2022).